\documentclass[12pt]{article}
\usepackage{epsfig}
\def\be{\begin{equation}}
\def\ee{\end{equation}}
\def\beq{\begin{equation}}
\def\eeq{\end{equation}}
\def\beqar{\begin{eqnarray}}
\def\eeqar{\end{eqnarray}}
\def\barr{\begin{array}}
\def\earr{\end{array}}

\def\and{\qquad {\rm and } \qquad}

\def\p{\partial}

\def\slp{p \hspace{-1ex}/}
\def\sleps{ \epsilon \hspace{-1ex}/}
\def\slk{k \hspace{-1ex}/}

\def\ie{ {\it i.e.} }

\def\sbar{ \overline{s} }
\def\thmin{\theta_0}
\def\cmin{\cos \theta_0}

\def\eebar{$e^+e^-~$}
\def\eegz{$e^+e^- \to \gamma Z$}
\def\ggz{$\gamma\gamma Z~$}
\def\gzz{$\gamma ZZ~$}
\begin{document}
\renewcommand{\thefootnote}{\fnsymbol{footnote}}
\begin{flushright}
IISc-CHEP-3/04 \\
UWThPh-2004-10
\end{flushright}
\vskip .3cm

\begin{center}{\Large \bf \boldmath
Transverse beam polarization and CP-violating triple-gauge-boson couplings in
$e^+e^- \to \gamma Z$}
\vskip 1cm
{B. Ananthanarayan$^a$, Saurabh D. Rindani$^b$, 
Ritesh K. Singh$^a$, A. Bartl$^c$}
\vskip .5cm

{\it $^a$Centre for High Energy Physics, 
Indian Institute of Science\\ Bangalore
560 012, India\\~ \\
$^b$Theory Group, Physical Research Laboratory\\ Navrangpura, Ahmedabad 380 009,
India\\~ \\
$^c$ Universit\" at Wien,
Institut fur Theoretische Physik\\
Boltzmanngasse 5,
A-1090 Vienna, Austria}
\end{center}
\vskip 1cm
\begin{quote}
\centerline{\bf Abstract}
We show that an anomalous CP-violating $\gamma\gamma Z $ vertex gives rise 
to a novel
asymmetry with transversely polarized electron and positron beams
in the process $e^+e^- \to \gamma Z$. This asymmetry, which is odd under naive
time reversal, is proportional to the real part of the $\gamma\gamma Z $
CP-violating coupling. This is in contrast to the simple forward-backward
asymmetry of the $\gamma$ (or $Z$) with unpolarized or
longitudinally polarized beams studied earlier, which is even under naive time
reversal, and is proportional to the imaginary part. We estimate 
the sensitivity of future
experiments to the determination of CP-odd $\gamma\gamma Z $ and $\gamma Z
Z $ couplings using these asymmetries and transversely polarized beams.

\end{quote}

\newpage
\section{Introduction}

A future linear \eebar collider operating at a centre-of-mass (cm) energy of
several hundred GeV would contribute greatly to a precise determination of the
parameters of known particles and their interactions, as well as to
the constraining of new physics. Longitudinal polarization of the $e^+$ and
$e^-$ beams, which is expected to be feasible at such colliders, would be
helpful in reducing background as well as enhancing the sensitivity. It has
been realized that spin rotators can be used to convert the longitudinal
polarizations of the beams to transverse polarizations. The question has
often been asked if such transverse polarization can be put to use to shed
light on interactions or parameter ranges not accessible with longitudinal
polarization, or to enhance their sensitivity. This question has not been
discussed exhaustively in the current context as yet, though there have been
some recent studies \cite{rizzo}-\cite{Ananthanarayan:2003wi}.

The role of transverse polarization in the context of CP violation has been
studied in \cite{Choi:2001ww}-\cite{Burgess:1990ba}.
Since transverse beam polarization provides an additional reference coordinate
axis in addition to the \eebar beam direction, there is the possibility of
studying the azimuthal distribution of a single final-state particle. This has
the advantage that the polarization of the produced particle, 
and hence its decay
distribution, need not be measured. In \cite{Ananthanarayan:2003wi} it was
pointed out that an azimuthal distribution of a final-state particle $A$ in a
semi-inclusive process $ e^+e^- \to A + X$ arising from the interference
between a standard model (SM) contribution and a new-physics contribution
arising at some high scale cannot contain a CP-violating part if the
new-physics contribution arises from chirality-conserving vector (V) or
axial-vector (A) type of interaction, neglecting the electron mass. This
result, with the SM contribution restricted to a virtual photon exchange, can
be deduced from the work of Dass and Ross \cite{Dass:1975mj}. In
\cite{Ananthanarayan:2003wi}, this was generalized to include virtual $Z$
exchange as well.  On the
other hand, chirality-violating scalar (S) and tensor (T) interactions can give
rise to a simple CP-odd azimuthal asymmetry, as for example, in $e^+e^- \to t
\overline{t}$ \cite{Ananthanarayan:2003wi}.

The above results were obtained with the condition that the SM contribution
arises only through $s$-channel exchange of virtual photon and $Z$. The
possibility of $t$- and $u$-channel exchange of an electron was not considered.
 Moreover, since the new physics is supposed to arise at a high scale, 
no $t$- or $u$-channel exchange of new particles was included. The results 
may get somewhat modified if these effects are taken into account.
In particular, the $t$- or $u$-channel exchange would introduce an extra
dependence on the scattering (polar) angle $\theta$. In a process where $A$ is
its own conjugate, there may be a consequent 
forward-backward asymmetry corresponding to $\theta \to \pi - \theta$, which is
CP odd. It is well-known that such an asymmetry could arise without
transverse polarization (see, for example, \cite{Czyz:1988yt,
Choudhury:1994nt, 
Rindani:1997qn}). However, such a forward-backward asymmetry, in the absence of
transverse polarization, is even under naive time reversal $T$ (\ie$\!\!$, 
reversal
of particle spins and momenta). Hence the CPT theorem implies that the
contribution comes only from an absorptive part in one of the interfering
amplitudes (see, for example, \cite{Rindani:1994ad}). 
Thus, such a symmetry is only sensitive to the imaginary parts of
the  new-physics couplings. 

In this paper we investigate the interesting possibility that if there is 
transverse polarization, a T-odd but CP-even azimuthal asymmetry can be
combined with the T-even but CP-odd forward-backward asymmetry to give an
asymmetry which is both CP odd as well as T odd. In this case, the CPT theorem
dictates that such an asymmetry meaure the real part of the new-physics
couplings. The process we have chosen is \eegz, where the final-state particles
are both self-conjugate\footnote{A similar asymmetry has been considered for
neutralino pair production in \cite{Choi:2001ww}.}.
This process occurs at tree level in SM.  A
CP-violating contribution can arise if anomalous CP-violating \ggz and \gzz 
couplings are present. The interference of the contributions from these
anomalous couplings with the SM contribution gives rise to the expected
polar-angle forward-backward asymmetry, as well as new combinations of polar
and azimuthal asymmetries. In particular, there is a CP-odd, T-odd asymmetry,
which is proportional to the real part of the \ggz coupling. This real part
cannot be probed without transverse polarization\footnote{An analogous
situation is studied in \cite{Diehl:2003qz}, where transverse beam
polarization allows one to probe certain CP-conserving triple gauge-boson 
couplings which
cannot be probed with longitudinal polarization.}. There is an
accidental cancellation of a similar contribution arising from real part of
the \gzz coupling. 

\section{The process \eegz}

We now describe the details of our work. The process considered is 
\begin{equation}
e^-(p_-,s_-)+e^+(p_+,s_+)\rightarrow \gamma (k_1)+Z(k_2).
        \label{process}
\end{equation}

The most general effective $CP$-violating Lagrangian  for $\gamma\gamma Z$
and $\gamma ZZ$
interactions, consistent with Lorentz invariance and electromagnetic gauge
invariance, and retaining terms upto dimension 6, can be written as
\be
\barr{rcl}
{\cal L} &= & \displaystyle
   e \frac{\lambda_1}{ 2 m_Z^2} F_{\mu\nu}
    \left( \p^\mu Z^\lambda \p_\lambda Z^\nu
          - \p^\nu Z^\lambda \p_\lambda Z^\mu
      \right)
       \\[2ex]
& & \displaystyle
      +\frac{e}{16 c_W s_W} \frac{\lambda_2}{m_Z^2}
       F_{\mu\nu}F^{\nu \lambda}
       \left(\p^\mu Z_\lambda + \p_\lambda Z^\mu   \right),
\earr
      \label{lagrangian}
\ee
where
$c_W=\cos \theta_W$ and $s_W=\sin \theta_W$ and $\theta_W$ is the
weak mixing angle. 
Terms involving divergences of the vector fields have been dropped from the
Lagrangian as they would not contribute when the corresponding particle is
on the mass shell, or is virtual, but coupled to a conserved fermionic current.
Since we will
 neglect the electron mass, the corresponding current can
be assumed to be conserved. 
We have not tried to impose full $SU(2)_L\times U(1)$
invariance, but only electromagnetic gauge invariance, as this is more general.

\begin{figure}[htb]
\centering
\epsfig{file=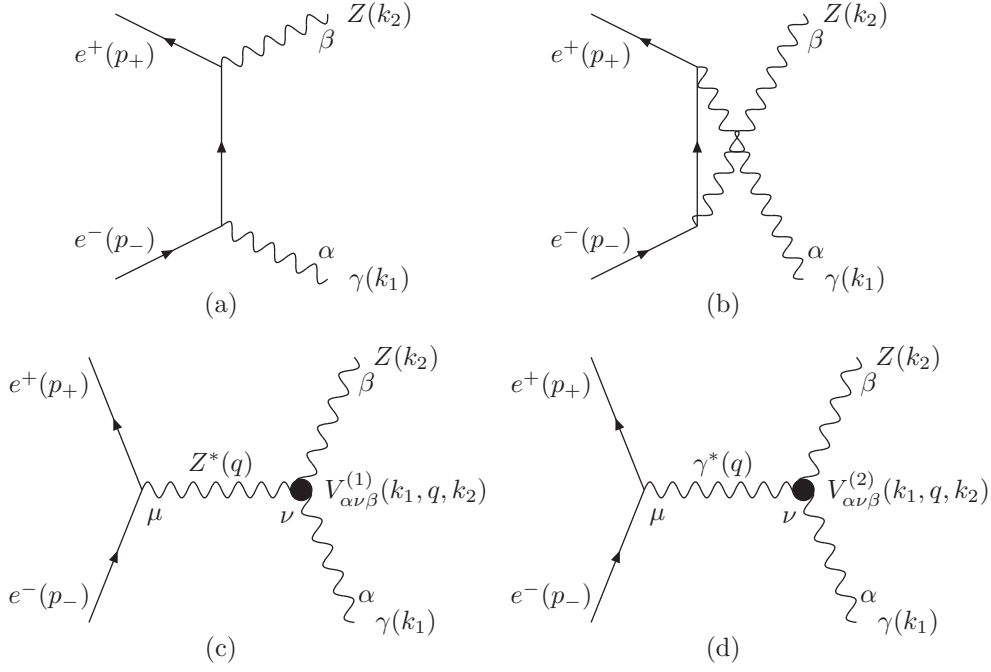}
\caption{Diagrams contributing to the process \eegz. Diagrams (a) and (b)
are SM contributions and diagrams (c) and (d) correspond to contributions from
the anomalous \gzz and \ggz couplings.}
\label{feynman}
\end{figure}

The SM diagrams contributing to the process (\ref{process}) are shown in Figs.
\ref{feynman} (a) and \ref{feynman} (b), which correspond to 
$t$-- and a $u$--channel electron exchange, while
the extra piece in the Lagrangian (\ref{lagrangian}) introduces two
$s$--channel diagrams with $\gamma$-- and $Z$--exchange respectively, shown in
Figs. \ref{feynman} (c) and \ref{feynman} (d).
 The corresponding matrix element is then given by
\begin{equation}
{\cal M}= {\cal M}_a +{\cal M}_b +{\cal M}_c +{\cal M}_d ,
       \label{amplitude}
\end{equation}
where
\be
\barr{rcl}
{\cal M}_a & = & \displaystyle
   \frac{e^2}{4c_Ws_W}\, \bar{v}(p_+)\: \sleps(k_2) (g_V - g_A\gamma_5)
   \frac{1}{\slp_- - \slk_1}\: \sleps(k_1)\: u(p_-),
       \\[2ex]
{\cal M}_b &=& \displaystyle \frac{e^2}{4c_W s_W}\, \bar{v}(p_+) \: \sleps(k_1)
  \frac{1}{\slp_- - \slk_2} \sleps(k_1) (g_V-g_A\gamma_5) u(p_-),
       \\[2ex]
{\cal M}_c & = & \displaystyle \frac{ie^2\lambda_1}{4c_W s_W m_Z^2}\,
       \bar{v}(p_+)\gamma _\mu (g_V-g_A\gamma_5)u(p_-)
   \frac{(-g^{\mu\nu} + q^\mu q^\nu / m_Z^2)}
        {q^2-m_Z^2}
  V^{(1)}_{\alpha\nu\beta}(k_1,q,k_2) \epsilon^\alpha(k_1)
        \epsilon^\beta(k_2),
     \\[2ex]
{\cal M}_d &= & \displaystyle
      \frac{ie^2\lambda_2}{4 c_W s_W m_Z^2}\,
   \bar{v}(p_+)\gamma _\mu u(p_-)
         \frac{(-g^{\mu\nu})}
              {q^2}
     V^{(2)}_{\alpha\nu\beta}(k_1,q,k_2)
      \epsilon^\alpha (k_1)\epsilon^\beta(k_2).
\earr
     \label{matrix elem}
\ee
We have used $q = k_1 + k_2$, and the tensors $V^{(1)}$ and $V^{(2)}$
corresponding to the three-vector vertices are given by
\be
\barr{rcl}
V^{(1)}_{\alpha\nu\beta}(k_1,q,k_2)
   &= &  \displaystyle
k_1\cdot q \: g_{\alpha\beta} \: k_{2\nu} + k_1\cdot k_2 g_{\alpha\nu} q_\beta
- k_{1\beta} \: q_\alpha \: k_{2\nu}  - k_{1\nu} \:q_\beta \: k_{2\alpha}
   \\[2ex]
V^{(2)}_{\alpha\nu\beta}(k_1,q,k_2)
   & = & \displaystyle
  \frac{1}{2}   \left[
  g_{\alpha\beta}
            \left( k_2\cdot q \: k_{1\nu} - k_1\cdot q \: k_{2\nu} \right)
        - g_{\nu\alpha}
            \left( k_2\cdot q \: k_{1\beta} + k_1 \cdot k_2 \: q_\beta \right)
      \right.
      \\[2ex]
   & &  \displaystyle \hspace{2em}
    \left.
       + g_{\nu\beta}
            \left( k_1\cdot k_2 \: q_\alpha - k_1\cdot q \: k_{2\alpha} \right)
        + q_\alpha \: k_{2\nu} \: k_{1\beta}
        + q_\beta \: k_{1\nu} \: k_{2\alpha}
    \right]
{}.
\earr
     \label{vertices}
\ee

In the above,
the vector and axial vector $Z$ couplings of the electron are given by
\begin{equation}
g_V = -1 + 4\sin^2\theta_W ;\quad g_A = -1
     \label{gVgA}.
\end{equation}

For compactness, we introduce the notation :
\be
\barr{rcl}
\sbar & \equiv & \displaystyle \frac{s}{m_Z^2}  \; ,
\\[2ex]
   {\cal B} & = & \displaystyle
      \frac{\alpha^2}{16 s_W^2 m_W^2 \sbar}
     \left( 1 - \frac{1}{\sbar}   \right)
     (g_V^2+g_A^2) \; ,
\\[2ex]
C_{A} & = & \displaystyle
        \frac{\sbar - 1}{4 (g_V^2+g_A^2)}\:\left\{           
    (g_V^2+g_A^2+
    (g_V^2-g_A^2)\, P_e P_{\overline{e}}\, \cos 2\phi
    ) \,
    {\rm Im} \lambda_1 
    \right. \\
    & & \left. 
                 - 
		 g_V (1+P_e P_{\overline{e}}\, \cos 2 \phi) \,
		 {\rm Im\lambda_2} 
		 -
		 g_A \, P_e P_{\overline{e}}\,
		 \sin 2\phi\, 
		 {\rm Re} \lambda_2 
		 \right\}
		 \;  .
   \label{notation}
\earr
\ee

Using eqns. (\ref{amplitude} - \ref{notation}),
we obtain the differential cross section
for the process (1) to be
\be
\displaystyle
\frac{d\sigma}{d\Omega    } =
    {\cal B}
\left[
       \frac{1}{\sin^2 \theta}
          \left( 1 + \cos^2 \theta + \frac{4 \sbar}{( \sbar - 1)^2}
	- P_e P_{\overline{e}}\frac{g_V^2-g_A^2}{g_V^2+g_A^2}
	\sin^2 \theta \cos 2\phi 
           \right)
     + C_A \cos \theta
\right]  \; ,
    \label{diff c.s.}
\ee
where $\theta$ is the angle between photon and the $e^-$ directions, and $\phi$
is the azimuthal angle of the photon, with $e^-$ direction chosen as the $z$
axis and the direction of its transverse polarization chosen as the $x$ axis.
The $e^+$ polarization direction is chosen parallel to the $e^-$ polarization
direction. $P_e$ and $P_{\overline{e}}$ are respectively the degrees of
polarization of the $e^-$ and $e^+$. 
We have kept only terms of leading order in the anomalous couplings, 
since they are
expected to be small. The above expression may be obtained either 
by using standard trace techniques for Dirac spinors with a
transverse spin four-vector, 
or by first calculating helicity amplitudes and then writing
transverse polarization states in terms of helicity states \cite{Hikasa:1985qi}. 

We will assume a cut-off  $\thmin$ on the polar angle
$\theta$ of the photon
in the forward and backward directions. This cut-off is needed to stay away
from the beam  pipe. It can further be chosen to optimize the sensitivity.
The total cross section corresponding to the cut
$\thmin < \theta < \pi - \thmin$
can then be easily obtained by integrating the differential cross section
above.

It is interesting to note that the contribution of the interference between 
the SM amplitude and the anomalous amplitude vanishes for $s=m_Z^2$. The reason
for this is that for $s=m_Z^2$ the photon in the final state is produced with
zero energy and momentum. As can be seen from eq. (\ref{vertices}), the
anomalous couplings vanish for $k_1=0$, leading to a vanishing intereference
term.

In order to understand the CP properties of various terms in the differential
cross section, we note the following relations:

\beq\label{ctheta}
\vec{P}\cdot \vec{k}_1 =\frac{\sqrt{s}}{2} \vert \vec{k}_1 \vert \cos\theta\;,
\eeq
\beq\label{s2ts2p}
(\vec{P} \times \vec{s}_- \cdot \vec{k}_1)( \vec{s}_+\cdot \vec{k}_1) + 
(\vec{P} \times \vec{s}_+ \cdot \vec{k}_1) (\vec{s}_-\cdot \vec{k}_1) 
= \frac{\sqrt{s}}{2} \vert \vec{k}_1 \vert^2 \sin^2\theta \sin 2\phi\; ,
\eeq
\beq\label{s2tc2p}
(\vec{s}_- \cdot \vec{s}_+) (\vec{P}\cdot \vec{P} \vec{k}_1 \cdot \vec{k}_1 - 
\vec{P}\cdot\vec{k}_1 \vec{P}\cdot\vec{k}_1) - 2 (\vec{P}\cdot \vec{P}) 
( \vec{s}_-
\cdot \vec{k}_1) ( \vec{s}_+\cdot \vec{k_1})
 = \frac{s}{4} \vert \vec{k}_1 \vert^2 
\sin^2\theta \cos 2\phi\; .
\eeq 
where $\vec{P}=\frac{1}{2}(\vec{p}_- - \vec{p}_+)$.
Observing that the vector $\vec{P}$ is C and P odd, that the photon
momentum $\vec{k}_1$ is C even but P odd, and that the spin vectors
$\vec{s}_{\pm}$ are P even, and go into each other under C, 
we can immediately check that only the left-hand side (lhs)
 of eq. (\ref{ctheta}) is
CP odd, while the lhs of eqs. (\ref{s2ts2p}) and (\ref{s2tc2p})
are CP even. Of all the above, only the lhs of (\ref{s2ts2p}) is odd under
naive time reversal T. 

We now define the following CP-odd asymmetries, which combine a
forward-backward asymmetry with an appropriate asymmetry in $\phi$, so as to
isolate appropriate anomalous couplings:
\begin{eqnarray}& \displaystyle A_1=
{1\over \sigma_0}
\sum_{n=0}^3 (-1)^n
\left(
\int_{0}^{\cos \theta_0} d \cos\theta
\int_{\pi n/ 2}^{\pi(n+1)/  2} d\phi \,
{d \sigma \over d \Omega} -
\int_{-\cos \theta_0}^{0} d \cos\theta
 \int_{\pi n/ 2}^{\pi(n+1)/  2} d\phi \,
{d \sigma \over d \Omega} \right)  
\end{eqnarray}
\begin{eqnarray}& \displaystyle A_2=
{1\over \sigma_0}
\sum_{n=0}^3(-1)^n \left(
\int_{0}^{\cos \theta_0} d \cos\theta
 \int_{\pi(2 n-1)/4}^{\pi(2 n+1)/4} d\phi \,
{d \sigma \over  d \Omega} -
\int_{-\cos \theta_0}^0 d \cos\theta
 \int_{\pi (2 n-1)/4}^{\pi(2 n+1)/4} d\phi \,
{d \sigma \over  d \Omega} \right) &
\end{eqnarray}
\begin{eqnarray}& \displaystyle A_3=
{2\over \sigma_0}\left\{
\int^{\cos \theta_0}_{0} d \cos\theta \left(
\int_{-\pi/4}^{\pi/4} d\phi \,
{d \sigma \over d \Omega} +
\int_{3 \pi /4}^{5 \pi/4} d\phi \,
{d \sigma \over d \Omega} \right)  \right. & \nonumber\\
& \displaystyle \left.-
\int_{-\cos \theta_0}^{0} d \cos\theta \left(
\int_{-\pi/4}^{\pi/4} d\phi \,
{d \sigma \over d \Omega} +
\int_{3 \pi /4}^{5 \pi/4} d\phi \,
{d \sigma \over d \Omega} \right) \right\} &
\end{eqnarray}
with
\begin{eqnarray}
& \displaystyle \sigma_0 \equiv \sigma_0(\thmin)=
\int_{-\cos \theta_0}^{\cos \theta_0} d \cos\theta
\int_{0}^{2 \pi} d\phi \,
{d \sigma \over d \Omega            }\, . &
\end{eqnarray}

These are easily evaluated to be
\begin{eqnarray}
&   \displaystyle A_1(\theta_0)=              
       - {\cal B}' 
	\, g_A\,  P_e P_{\overline{e}}\, {\rm Re} \lambda_2
		 \; &  ,
		 \end{eqnarray}
\begin{eqnarray}
& \displaystyle A_2(\theta_0)=
        {\cal B}'\,
    P_e P_{\overline{e}}\, ((g_V^2-g_A^2)\, 
    {\rm Im} \lambda_1 - g_V \, {\rm Im}\lambda_2)  &
    \end{eqnarray}
\begin{equation}
 \displaystyle A_3(\theta_0)=
        {\cal B}'\,
\left[ \frac{ \pi}{2} (  (g_V^2+g_A^2)\, {\rm Im} \lambda_1\! -\!g_V\, 
{\rm Im}\lambda_2)
+ P_e P_{\overline{e}}
((g_V^2-g_A^2)
  \,  {\rm Im} \lambda_1\! -\! g_V \, {\rm Im}\lambda_2)\right].  
    \end{equation}
\begin{eqnarray}
& \displaystyle \sigma_0=4 \pi {\cal B}
  \left[ \left\{ \frac{\sbar^2 + 1}{(\sbar - 1)^2}
                   \ln \left( \frac{1 + \cmin }
                                   {1 - \cmin }
                            \right)
                 - \cmin
          \right\}\right]\; .
\end{eqnarray}
In the above equations, we have defined
\begin{equation}
{\cal B}' =
\frac{{\cal B} (\sbar - 1)\cos^2\theta_0}  
	{ (g_V^2+g_A^2)\sigma_0(\theta_0)}.
\end{equation}

\begin{figure}[ht!]
\centering
\epsfig{file=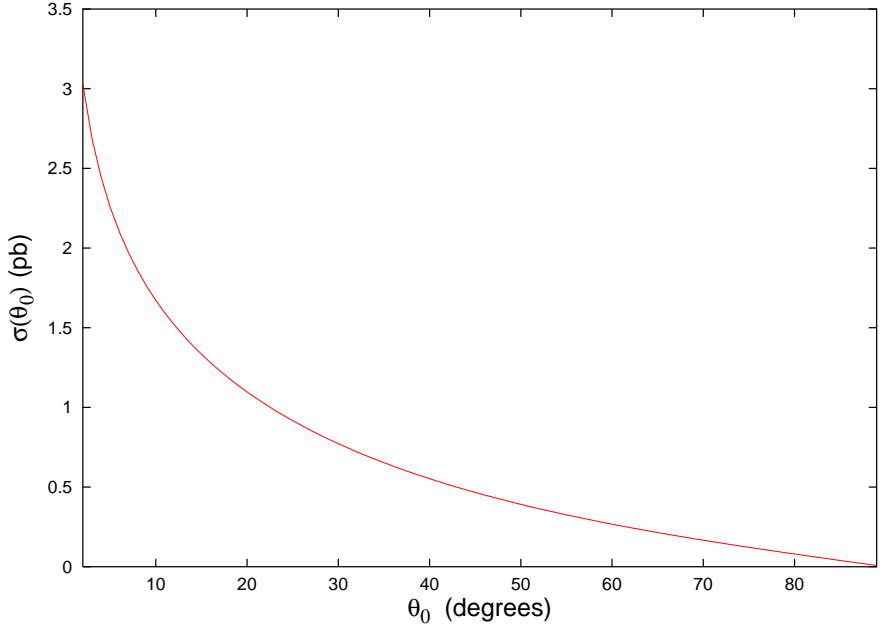,angle=-90,width=12cm}
\caption{The SM cross section with a cut-off $\theta_0$ in the forward and 
backward directions plotted as a function of $\theta_0$.}
\label{cs}
\end{figure}
\begin{figure}[htb]
\centering
\epsfig{file=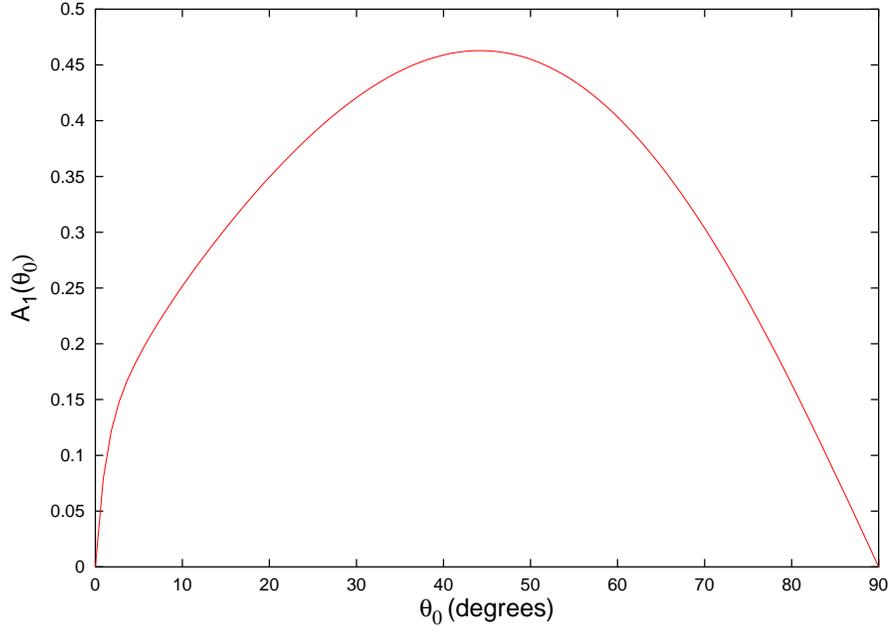,angle=-90,width=12cm}
\caption{The asymmetry $A_1(\theta_0)$ defined in the text plotted as a function
of the cut-off $\theta_0$ for a value of Re $\lambda_2 = 1$.}
\label{asym1}
\end{figure}
\begin{figure}[ht!]
\centering
\epsfig{file=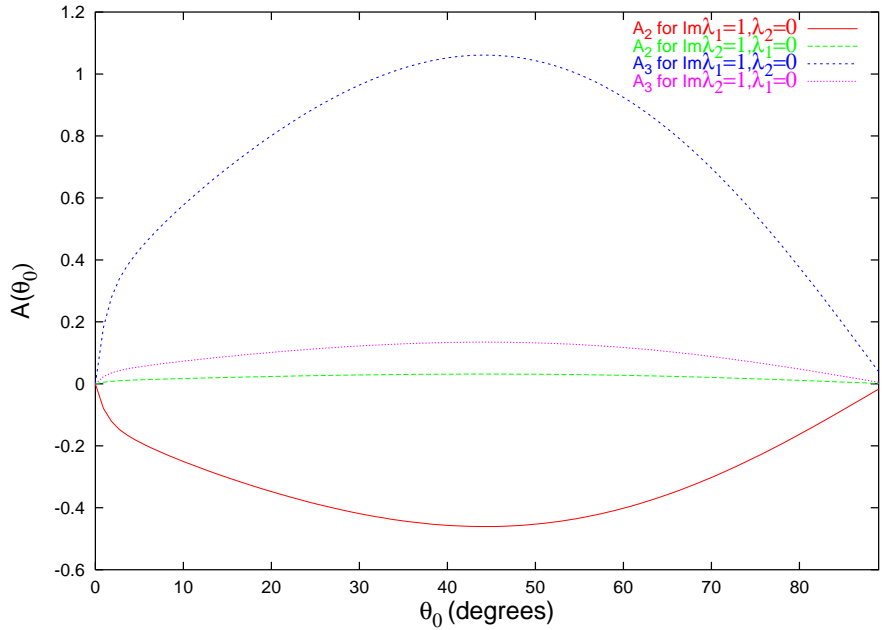,angle=-90,width=12cm}
\caption{The asymmetries $A_2(\theta_0)$ and $A_3(\theta_0)$ defined in the text
plotted as a function of the cut-off $\theta_0$ for values Im $\lambda_1 = 1$,
Im $\lambda_2 = 0 $, and Im $\lambda_1 = 0$, Im $\lambda_2 = 1 $.}
\label{asym2}
\end{figure}
\begin{figure}[ht!]
\centering
\epsfig{file=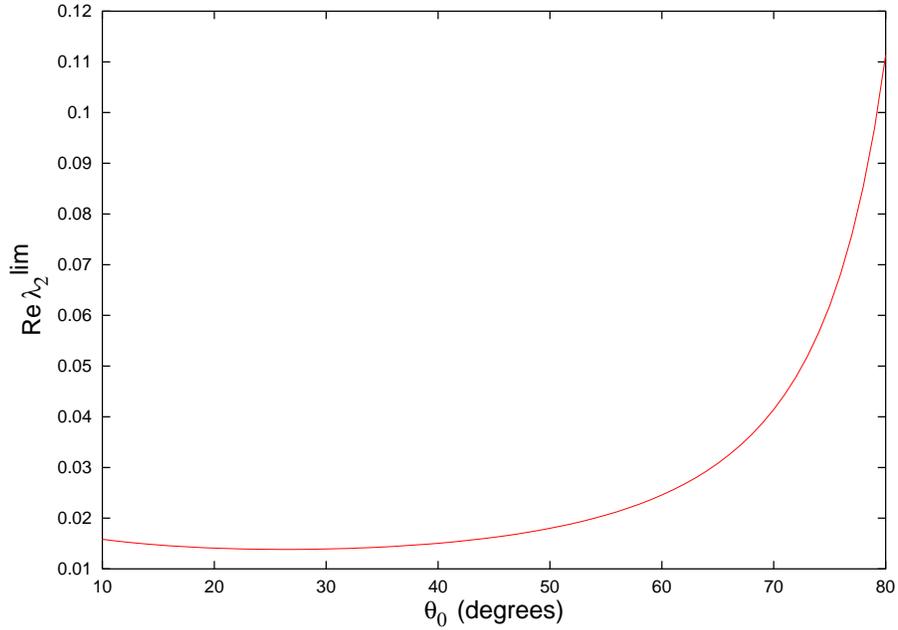,angle=-90,width=12cm}
\caption{The 90\% C.L. limit on Re $\lambda_2$ from the asymmetry $A_1(\theta_0)$
plotted as a funtion of the cut-off $\theta_0$.}
\label{a1lim}
\end{figure}
\begin{figure}[ht!]
\centering
\epsfig{file=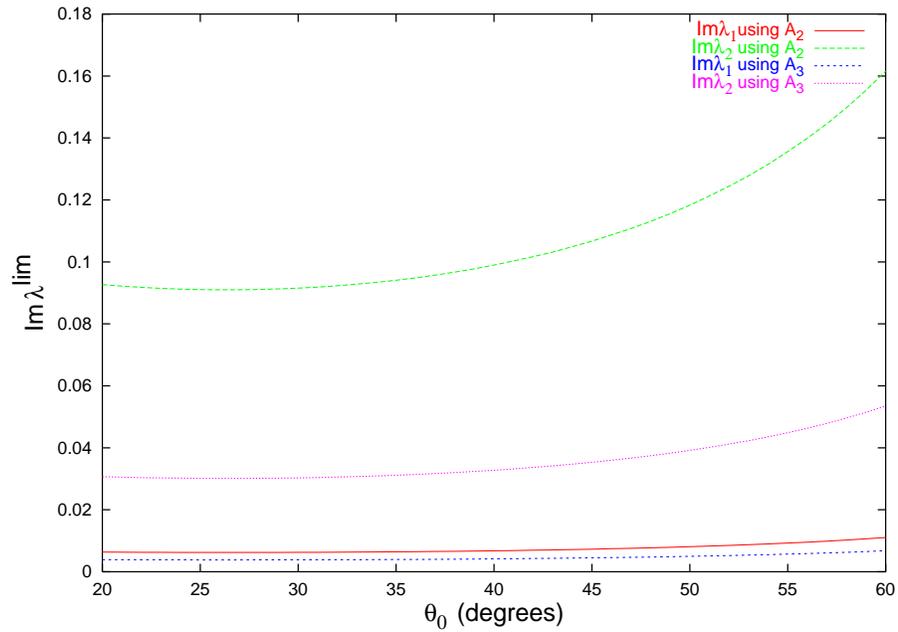,angle=-90,width=12cm}
\caption{The 90\% C.L. limits on Im $\lambda_1$ and Im $\lambda_2$, taken
nonzero one at a time, from the asymmetries $A_2(\theta_0)$ and $A_3(\theta_0)$,
plotted as a functions
of the cut-off $\theta_0$.}
\label{a2lim}
\end{figure}
\begin{figure}[ht!]
\centering
\epsfig{file=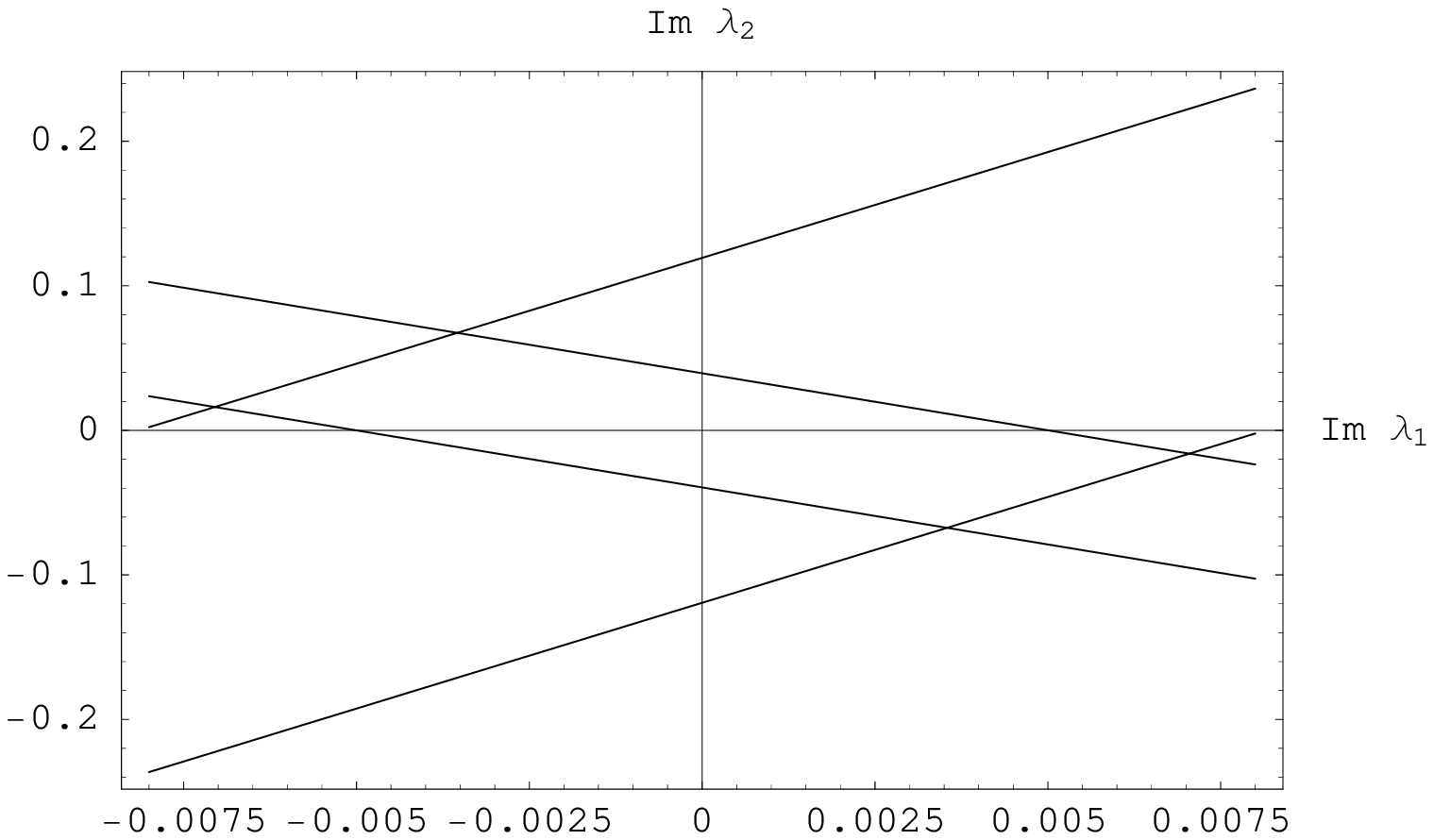,width=12cm}
\caption{90\% CL contours for the simultaneous determination of Im $\lambda_1$
and Im $\lambda_2$. The region inside the trapezium is the allowed region.}
\label{contour}
\end{figure}

We now make some observations on the above expressions which justify the 
choice of our asymmetries and highlight the novel features of our work.
It can be seen that
$\displaystyle A_1(\theta_0)$ is proportional to ${\rm Re} \lambda_2$,
and the other two asymmetries depend on ${\rm Im} \lambda_1$ and $ {\rm
Im}\lambda_2$. Moreover, the latter two measured simultaneously can be used
to get limits on the two couplings ${\rm Im} \lambda_1$ and ${\rm
Im}\lambda_2$. It is interesting that $A_1$ does not depend on $\lambda_1$,
which is the result of an accidental cancellation. This would not be
the case, for example, if the $Z$ in the $s$-channel exchange in Fig. 1(a) were
different from the $Z$ produced in the final state, so that their couplings to
the the electron were different.

Note that the vector coupling $g_V$ of the electron is small. As a result, the
asymmetries $A_2$ and $A_3$ are relatively insensitive to ${\rm Im}\lambda_2$.
However, in $A_3$, there is a partial cancellation of the ${\rm Im} \lambda_1$
contribution, making $A_3$ more sensitive to ${\rm Im}\lambda_2$ than $A_2$.
This is borne out by our numerical results, see below.

\section{Numerical Results}

We now present our numerical results. The cross section with a cut-off $\theta_0$
on $\theta$ is plotted in Fig. \ref{cs} as a function of $\theta_0$.
Figs. \ref{asym1}, \ref{asym2} show the asymmetries as a function of the cut-off 
when the values of the anomalous couplings are taken to be nonzero one at a time.
All the asymmetries vanish not only for $\theta_0=0$, by definition, but also
for $\theta_0=90^{\circ}$, because they are proportional to $\cos\theta_0$.
They peak  at around $45^{\circ}$. 

We have 
calculated 90\% CL limits that can be obtained with a linear collider with
$\sqrt{s} = 500$ GeV, $\int L dt = 500$ fb$^{-1}$, $P_e = 0.8$, and 
$P_{\overline{e}} = 0.6$ making use of the asymmetries $A_i$. 
The limiting value $\lambda^{\rm lim}$ (\ie the respective real or imaginary
part of the coupling) is related to the value $A$ of the asymmetry for unit
value of the coupling constant by
\beq
\lambda^{\rm lim} = \frac{1.64}{A\sqrt{N_{SM}}},
\eeq
where $N_{SM}$ is the number of SM events.

$A_1$ depends on Re $\lambda_2$ alone, and can therefore place an 
independent limit on Re $\lambda_2$. We emphasize once again that
information on Re $\lambda_2$ cannot be obtained without transverse
polarization. 

Fig. \ref{a1lim} shows the 90\% CL limit on Re $\lambda_2$ as a
function of the cut-off. The asymmetries $A_2$ and $A_3$ depend on both Im
$\lambda_1$ and Im $\lambda_2$. Fig. \ref{a2lim} shows the 90\% CL limits on Im
$\lambda_i$ taken to be nonzero one at a time, using the asymmetries $A_2$ and
$A_3$. It can be seen from these figures that the limits are relatively
insensitive to the cut-off at least for small values of the cut-off. We find
that the best
limits are obtained for $\theta_0=26^{\circ}$, though any nearby value of
$\theta_0$ would give very similar results. These correspond to 
 Re $\lambda_2 = 0.0138$ (from $A_1$), Im $\lambda_1 = 0.00622$ (from $A_2$), 
Im $\lambda_1 = 0.00382$ (from $A_3$), 
Im $\lambda_2 = 0.0910$ (from $A_2$), and Im $\lambda_2 =
0.0301$ (from $A_3$). 

As stated earlier, because of $g_V$ being numerically small, the limits on 
Im $\lambda_2$, which appears in the expressions for the asymmetries multiplied
by $g_V$,  are worse than those on Im $\lambda_1$. However, it can also be seen
that $A_3$ fares better than $A_2$ so far as Im $\lambda_2$ is concerned.

Finally, we have also evaluated the simultaneous 90\% CL 
limits that can be obtained on Im
$\lambda_1$ and Im $\lambda_2$ by measurement of $A_2$ and $A_3$. For this we
have chosen $\theta_0=26^{\circ}$. The
corresponding contour for allowed values of the couplings for a null
result of the
measurement of $A_2$ and $A_3$ is shown in Fig. \ref{contour}. 
This contour is obtained by equating the asymmetry obtained simultaneously from
nonzero Im $\lambda_1$ as well as nonzero Im $\lambda_2$ to
$2.15/\sqrt{N_{SM}}$.
It can be seen that the simultaneous limits that can be obtained are weaker than
individual limits,
with numerical values Im $\lambda_1 = 0.00705$ and Im $\lambda_2 = 0.0674$.
The best limits are summarized in Table 1.
\begin{table}
\centering
\begin{tabular}{l|c|c|c|c}
\hline
Coupling & \multicolumn{3}{|c|} {Individual limit from}
 & Simultaneous limits \\
& $A_1$ & $A_2$ & $A_3$ &\\ 
\hline 
Re $\lambda_2$ & $1.38\times 10^{-2}$ & & & \\
Im $\lambda_1$ && $6.22\times 10^{-3}$ &$ 3.82\times 10^{-3}$ & $7.05\times
10^{-3}$ \\
Im $\lambda_2$ && $9.10\times 10^{-2}$ &$ 3.01\times 10^{-2}$ & $6.74\times
10^{-2}$ \\
\hline
\end{tabular}
\caption{90 \% CL limits on the couplings from asymmetries $A_i$ for a cut-off
angle of 26$^{\circ}$, $\sqrt{s} = 500$ GeV, and integrated luminosity of 500
fb$^{-1}$. The electron and positron transverse polarizations are assumed to be
respectively 0.8 and 0.6.}
\end{table}

\section{Conclusions}

In conclusion, 
we have studied a novel CP-violating asymmetry ($A_1$) in \eegz with anomalous
neutral gauge boson couplings, 
which is special to a neutral
final state, the observation of which needs both electron and positron 
transverse polarizations. This is of special interest in the context of 
the negative result
stated in \cite{Ananthanarayan:2003wi}, that the observation of CP violation in
a two-particle final state, without measuring the polarization of the
final-state particles, is not possible with transversely polarized beams, unless
there are chirality-violating couplings of the electron and positron. That
result depended on an analysis where $t$- and $u$-channel
particle exchanges were not taken into account. 

Forward-backward asymmetry of a neutral particle with unpolarized or
longitudinally polarized beams as a signal of CP violation has
been studied before. However,  the CPT theorem implies that in such a case the
asymmetry is proportional to the absorptive part of the amplitude. The asymmetry
$A_1$ that we study in the presence of transverse polarizations includes also an
azimuthal angle asymmetry, which makes it odd under naive time reversal. It is
thus proportional to the real part of the anomalous coupling. This real part
cannot be studied without transverse polarization. 

We have also made a numerical study of the limits on various couplings that
could be obtained at a future linear collider with $\sqrt{s}=500$ GeV and an
integrated luminosity of 500 fb$^{-1}$ assuming realistic transverse
polarizations of 80\% and 60\% respectively for $e^-$ and $e^+$, respectively.
The best limits are summarized in Table 1. We thus see that transverse
polarization would provide a sensitive test of anomalous couplings,
particularly, Re $\lambda_2$.
\vskip .2cm
\noindent Acknowledgements: This work arose out of discussions that took place
at the 8th Workshop on High Energy Physics 
Phenomenology (WHEPP8), held at the
Indian Institute of Technology, Mumbai, January 5-16, 2004. It is our pleasure
to thank the organizers for their kind hospitality. 
BA thanks the
Department of Science and Technology, Government of India, for support and also
thanks S. Dattagupta and S. Gupta for comments.
AB has been
supported by the 'Fonds zur F\"orderung der wissenschaftlichen Forschung'
of Austria, FWF Project No. P16592-N02 and by the European Community's
Human Potential Programme under contract HPRN-CT-2000-00149.


\begin{thebibliography}{99}

\bibitem{rizzo}
T.~G.~Rizzo,
JHEP {\bf 0302} (2003) 008
[arXiv:hep-ph/0211374];
T.~G.~Rizzo,
JHEP {\bf 0308} (2003) 051
[arXiv:hep-ph/0306283].

\bibitem{desch}
K.~Desch, LC Note LC-PHSM-2003-002; 
J.~A.~Aguilar-Saavedra {\it et al.}  [ECFA/DESY LC Physics Working Group
                  Collaboration],
arXiv:hep-ph/0106315;
K. M\"onig and J.
Sekaric, Talk presented at ECFA LC Study,
2004;  G. Pasztor, Summary talk at the ECFA LC Study, 2004.

\bibitem{Diehl:2003qz}
M.~Diehl, O.~Nachtmann and F.~Nagel,
Eur.\ Phys.\ J.\ C {\bf 32} (2003) 17
[arXiv:hep-ph/0306247].

\bibitem{Fleischer:1993ix}
J.~Fleischer, K.~Kolodziej and F.~Jegerlehner,
Phys.\ Rev.\ D {\bf 49} (1994) 2174.

\bibitem{Choi:2000hb}
S.~Y.~Choi, M.~Guchait, J.~Kalinowski and P.~M.~Zerwas,
Phys.\ Lett.\ B {\bf 479} (2000) 235
[arXiv:hep-ph/0001175].

\bibitem{Choi:2001ww}
S.~Y.~Choi, J.~Kalinowski, G.~Moortgat-Pick and P.~M.~Zerwas,
Eur.\ Phys.\ J.\ C {\bf 22} (2001) 563
[Addendum-ibid.\ C {\bf 23} (2002) 769]
[arXiv:hep-ph/0108117].

\bibitem{Ananthanarayan:2003wi}
B.~Ananthanarayan and S.~D.~Rindani,
arXiv:hep-ph/0309260;
LC Note LC-TH-2003-099

\bibitem{Couture:1992if}
G.~Couture,
Phys.\ Lett.\ B {\bf 305} (1993) 306.

\bibitem{Stodolsky:1984ez}
L.~Stodolsky,
Phys.\ Lett.\ B {\bf 150} (1985) 221.

\bibitem{Burgess:1990ba}
C.~P.~Burgess and J.~A.~Robinson,
Int.\ J.\ Mod.\ Phys.\ A {\bf 6} (1991) 2707.

\bibitem{Dass:1975mj}
G.~V.~Dass and G.~G.~Ross,
Phys.\ Lett.\ B {\bf 57} (1975) 173;
Nucl.\ Phys.\ B {\bf 118} (1977) 284.

\bibitem{Czyz:1988yt}
H.~Czyz, K.~Kolodziej and M.~Zralek,
Z.\ Phys.\ C {\bf 43} (1989) 97.

\bibitem{Choudhury:1994nt}
D.~Choudhury and S.~D.~Rindani,
Phys.\ Lett.\ B {\bf 335} (1994) 198
[arXiv:hep-ph/9405242].

\bibitem{Rindani:1997qn}
S.~D.~Rindani and J.~P.~Singh,
Phys.\ Lett.\ B {\bf 419} (1998) 357
[arXiv:hep-ph/9703380].

\bibitem{Rindani:1994ad}
S.~D.~Rindani,
Pramana {\bf 45} (1995) S263
[arXiv:hep-ph/9411398].

\bibitem{Hikasa:1985qi}
K.~I.~Hikasa,
Phys.\ Rev.\ D {\bf 33} (1986) 3203.


\end{thebibliography}
\end{document}